\documentclass[twocolumn]{elsart}
\def\ApJ{{\it Astrophys. J}}
\def\MNRAS{{\it Mon. Not. R. Astron. Soc}}

\def\ApSS{{\it Astrophysics and Space Science}}

\usepackage{natbib}
\begin{document}
\runauthor{Fan and Piran}
\begin{frontmatter}
\title{A Canonical High Energy Afterglow Emission Light Curve?
}
\author[HU,PMO]{Yi-Zhong Fan
}
\author[HU]{Tsvi Piran\thanksref{Someone}}


\address[HU]{Racah Inst. of Physics, Hebrew University, Jerusalem 91904, Israel}
\address[PMO]{Purple Mountain Observatory, Chinese Academy of Science,
Nanjing 210008, China}
\thanks[Someone]{Supported by the US-Israel BSF.}
\begin{abstract}
We present self consistent calculations of Synchrotron self Compton
(SSC) radiation that takes place within the afterglow blast wave and
External inverse Compton (EIC) radiation that takes place when flare
photons (produced by an internal process) pass through the blast
wave. We show that if our current interpretations of the Swift XRT
data are correct, there should be a canonical high energy afterglow
emission light curve.  We expect that GRBs with a long term X-ray
flattening or X-ray flares should show similar high energy features.
The EIC emission, however, is long lasting and weak and might be
outshined by the SSC emission of the forward shock. The high energy
emission could be well detected by the soon to be launched  GLAST
satellite. Its detection could shed new light on the conditions
within the emitting regions of GRBs.
\end{abstract}
\end{frontmatter}

\section{Introduction}
\label{sec:SSC1} Very high energy emission provides us with another
window on the condition within the emitting region in Gamma-ray
Bursts (GRBs). Such a window is very important in view of the
present confusion between different modifications to the standard
afterglow model proposed to explain the recent observation of {\it
Swift}. The upcoming high energy observatory GLAST is an ideal tool
to detect such emission. Together with the {\it Swift} X-ray
Telescope (XRT), it would provide a very wide band monitoring of the
afterglow that might enable us to distinguish between the different
models. We discuss here several predictions of current models for
the expected high energy light curves and their relation to the
X-ray afterglow.

\section{SSC emission of the forward shock}
\begin{figure}
\begin{picture}(0,150)
\put(0,0){\includegraphics{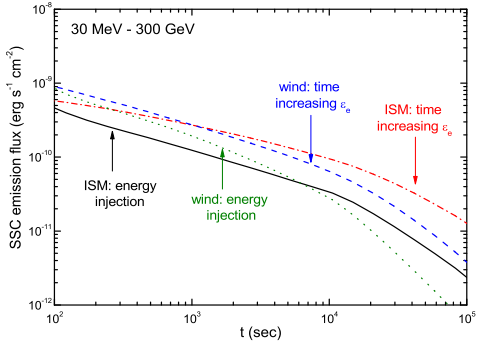}}
\end{picture}
\caption{The SSC radiation of the forward shock: the energy
injection case and the time increasing $\epsilon_e$ (Fan et al.
2007).} \label{fig:SSC}
\end{figure}
The dominant source of long lasting high energy GRB afterglow
emission is SSC of the hot electrons in the forward external
shock. At early stages of the afterglow  when the cooling of most
electrons is important, the luminosity of the SSC emission,
$L_{_{\rm SSC}}$, is related to the luminosity of the synchrotron
radiation, $L_{_{\rm syn}}$,  as $L_{_{\rm SSC}} \sim {Y} L_{\rm
syn}$, where ${Y}$ is the Compton parameter. The X-ray luminosity
$L_{X}$, is a small fraction of $L_{_{\rm syn}}$. We would  like
to use it as a proxy for the total luminosity and we  define a
factor $\epsilon_{\rm X} \equiv L_{X}/L_{\rm syn}$ so that
$L_{_{\rm SSC}} \sim {Y} L_{X}/\epsilon_{\rm X}$. A wide band SSC
afterglow data and the X-ray data will have quite similar temporal
behaviors, as long as $\epsilon_{\rm X}$ does not vary
significantly with time. Overall we expect, therefore, that $L_X$
and $L_{SSC}$ should be highly correlated. In other words, the
GRBs with slowly decaying X-ray light curve should show similar
high energy feature. This is confirmed by the more detailed
analysis (Fan et al. 2007; Wei \& Fan 2007) and by the numerical
results, as shown in Fig.\ref{fig:SSC}.

Energy injection and time increasing $\epsilon_e$, have been
proposed to explain the slowly declining X-ray phase seen in many
GRB afterglows. The resulting SSC light curves in both cases follow
the X-ray. Although the SSC light curves are somewhat different the
difference is probably too small to distinguish between the two
modifications (Fan et al. 2007). However, the typical SSC frequency
$\nu_m^{SSC}\propto \epsilon_e^4 E_k$, where $E_k$ is the total
energy of the outflow. The strong $\epsilon_e$ dependence  suggests
a significant difference in the time evolution of the high energy
spectrum between the two models.

\section{Possible high energy emission associated with flares}
X-ray flares during the afterglow have been detected by BeppoSAX
(Piro et al. 2005) and confirmed by {\it Swift} to exist in a
large fraction of the afterglows (Nousek et al. 2006; Zhang et al.
2006). Such flares should be accompanied by a high energy emission
either because of SSC emission of the electrons powering the
flares or because  EIC upscattering that takes place when flare
photons pass through a hot blast wave.

{\bf GeV flash.} We consider first the direct SSC emission
associated with X-ray flares. The typical frequency of the
upscattered X-ray photons depends on the Lorentz factor of the
scattering electrons. The magnetic energy density ($B$) at a radius
$R_{\rm flare}$ can be estimated as: $ B \sim 250~{\rm
Gauss}~\varepsilon^{1/2} L_{\rm x,49}^{1/2}\Gamma^{-1}R_{\rm
flare,17}^{-1}$, where $\varepsilon \equiv \epsilon_B/\epsilon_e$.
For this value of the magnetic field the peak energy of the flare
photons $E_{\rm p} \sim 0.2$ keV  requires a typical random Lorentz
factor of the emitting electrons: $\gamma_{\rm e,m} \sim
800~\varepsilon^{-1/4}L_{X,49}^{-1/4} R_{\rm flare,15}^{1/2}(E_{\rm
p}/0.2~{\rm keV})^{1/2}$. With this Lorentz factor the expected SSC
emission peaks at
\begin{eqnarray} \nu_p^{\rm
ssc} &\sim& 0.3 {\rm GeV}~\varepsilon^{-1/2}L_{X,49}^{-1/2} R_{\rm
flare,15}\nonumber\\
&& (E_{\rm p}/0.2~{\rm keV})^{2}.
\end{eqnarray}
The total fluence of the SSC emission of the flare shock is
comparable to that of the X-ray emission, typically
$10^{-7}\sim10^{-6}~{\rm erg~cm^{-2}}$. In a late internal shock
with $R_{\rm flare}\sim 10^{15}$ cm (Fan \& Wei 2005), a GeV flash
accompanying the X-ray flare is possible (Wei et al. 2006, however
see Wang et al. 2006). In an external shock, a GeV-TeV flash is
predicted (see also Galli \& Piro 2007).

{\bf Extended EIC emission.} A second source of high energy emission
arises when X-ray flare photons that are produced by internal energy
dissipation are be inverse Compton upscattered by the external
shock's hot electrons. A central ingredient of this scenario is that
in the rest frame of the blast wave, the seed photons are highly
beamed. We take care of this effect, following the analysis of
Aharonian \& Atoyan (1981).

If the EIC emission duration is comparable to that of the X-ray
flare, the EIC luminosity can be estimated by
\begin{equation}
L_{_{\rm eln}} \sim 10^{49}~{\rm erg~s^{-1}}~ \epsilon_{e,-1}
E_{k,53} t_3^{-1}. \label{eq:L_electron}
\end{equation}
In the rest frame of the shocked material, the EIC emission peaks at
$\theta_{\rm sc}=\pi$ and it vanishes for small scattering angles.
This effect lowers the high energy flux significantly in two ways.
First, a fraction of the total energy is emitted out of our line of
sight and thus the received power is depressed (relative to the
isotropic seed photon case). Second, the strongest emission is from
$\theta \sim 1/\Gamma$ (Fan \& Piran 2006). Thus,  the high energy
EIC emission will be delayed by
\begin{equation}
T_{\rm p} \sim (4-k)t_f, \label{eq:t_p}
\end{equation}
after the flare (emitted at $t_f$). As  $T_p$ is much longer than
$\Delta T$, the duration of the soft X-ray flare the EIC high energy
flux would be low:
\begin{equation}
L_{_{\rm EIC}}\sim {L_{\rm eln}\over (T_{\rm p}/\Delta T)}.
\end{equation}

A comparison of the EIC high energy component with the SSC emission
from the forward shock my render the EIC high energy component
undetectable. At the time of the flare, 100-1000sec after the burst,
the forward shock emission peaks in far-UV to soft X-ray band. The
corresponding SSC a luminosity of the forward shock  around $t_{\rm
f}$ is $L_{_{\rm SSC}}\sim L_{_{\rm eln}}Y/(1+Y)$. This is
significantly larger than $L_{_{\rm EIC}}$ and the wide EIC flare
would be undetectable (cf. Wang et al. 2006).

\begin{figure}
\begin{picture}(0,150)
\put(0,0){\includegraphics{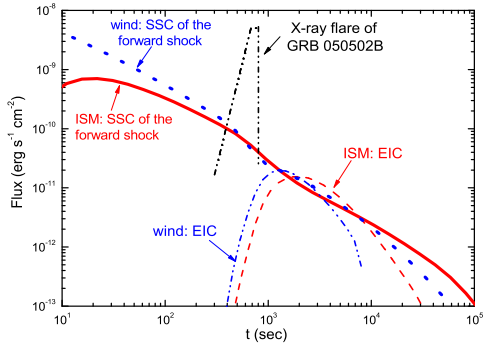}}
\end{picture}
\caption{The 30 MeV$-$300 GeV emission arising from flare photon
and a forward shock interaction for ISM/wind case (Fan et al.
2007). } \label{fig:EIC_flux}
\end{figure}

In special cases the EIC component might still be detectable. This
happens for bursts having a weak SSC emission and in which the
forward shock electrons are in slow cooling (before the X-ray flare
phase). The energy of these electrons will be lost mainly in the EIC
process and in this case the EIC luminosity will be enhanced.  If
the EIC emission dominates over the SSC emission, the high energy
light curve will flatten, as shown in Fig.\ref{fig:EIC_flux}. Such a
flattening could also arise by energy injection or due to an
increasing $\epsilon_e$. However, as shown in last section, in these
two scenarios, the X-ray and the high energy emission behaviors are
quite similar and flattening should be apparent also in the X-ray
signal. The EIC emission should, on the other hand show an X-ray
flare preceding high energy emission and not accompanying flat X-ray
light curve.

A numerical example illustrating the possible EIC emission
following the X-ray flare in GRB 050502B is shown in
Fig.\ref{fig:EIC_flux}.

\section{Summary and Discussion}
\begin{figure}
\begin{picture}(0,180)
\put(0,0){\includegraphics{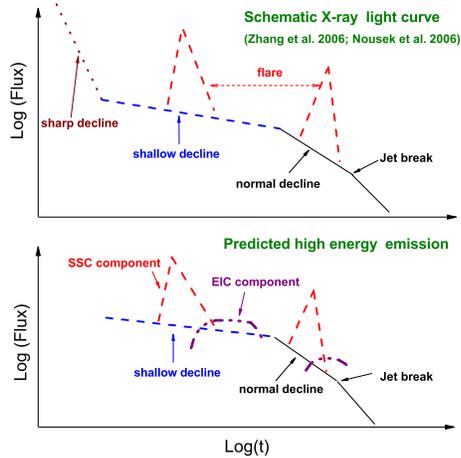}}
\end{picture}
\caption{The expected high energy afterglow signatures (the lower
panel), corresponding to the schematic X-ray afterglow light curve
based on {\it Swift} XRT data (Fan et al. 2007). }
\label{fig:summary}
\end{figure}
We have shown that if the current interpretation of the {\it
Swift} XRT data (the upper panel of Fig.\ref{fig:summary}) is
correct there should be a canonical high energy afterglow light
curve (see the lower panel of Fig.\ref{fig:summary} for
illustration). A detection of such a  high energy component  will
enable us to test current models of GRBs and their afterglow. A
high energy component that follows the lower energy light curve
will confirm that the low energy component is Synchrotron. If the
lower component is produced via Inverse Compton the Klein-Nishina
suppression will prevent a second upscattering. A detailed
comparison of the high energy and the low energy light curves, in
particular during the shallow decline phase, might even enable us
to distinguish between different modifications of the standard
afterglow model. This is because for the two most widely
considered models, the energy injection and time increasing
$\epsilon_e$, the time evolution of the high energy spectra are
very different. A long lasting high energy component that follows
a low energy flare would prove the internal origin of this flare
and the EIC model for the origin of the high energy emission. The
upcoming high energy observatory GLAST can thus play a key role in
exploring the GRB afterglow physics.

\end{document}